\documentclass[12pt]{article}
\usepackage{graphicx}
\usepackage{amsmath}
\usepackage{amsbsy}
\usepackage{amsfonts}
\usepackage{amsthm}
\begin{document}

\title{EYM equations in the presence of q-stars is scalar-tensor gravitational theories}

\author{Athanasios Prikas}

\date{}

\maketitle

Physics Department, National Technical University, Zografou
Campus, 157 80 Athens, Greece.\footnote{e-mail:
aprikas@central.ntua.gr}

\begin{abstract}
We study Einstein-Yang-Mills equations in the presence of a
gravitating non-topological soliton field configuration consisted
of a Higgs doublet, in Brans-Dicke and general scalar-tensor
gravitational theories. The results of General Relativity are
reproduced in the $\omega_{\textrm{BD}},\omega_0\rightarrow\infty$
limit. The numerical solutions correspond to a soliton star with a
non-abelian gauge field. We study the effects of the coupling
constant, the frequency of the Higgs field and the Brans-Dicke
field on the soliton parameters.
\end{abstract}

PACS number(s): 11.27+d, 04.40.-b, 04.70.Bw, 04.20.Jb

\newpage

\section{Introduction}

Einstein-Yang-Mills (EYM) equations have been investigated in
several systems, \cite{EYMeqs-review}. Bartnik and McKinnon found
particle-like non-abelian solutions of the coupled EYM theory,
\cite{Bartnik-McKinnon1,Bartnik-McKinnon2}. The coupling of the
EYM system with a scalar field may lead to several theories. We
mention gravitating Skyrmions, \cite{EYMskyrmions}, black hole
solutions in dilaton, \cite{EYMdilaton1,EYMdilaton2}, and massive
dilaton and axion gravity, \cite{EYMmassivedilaton}, and other
field configurations in the EYM-Higgs theory with a Higgs doublet,
\cite{EYMhiggs-doublet,EYMhiggs-doublet2} or with a Higgs triplet,
\cite{EYMhiggs-triplet1,EYMhiggs-triplet2,EYMhiggs-triplet3}.

Non-topological solitons and soliton stars have a long presence in
modern physics, \cite{non-top-sol-review}. In the present work we
need the theory of q-balls and q-stars. Q-balls are
non-topological solitons in Lagrangians with a global $U(1)$
symmetry, \cite{qballs-initial}, or an $SU(3)$ or $SO(3)$
symmetry, \cite{qballs-nonabelian}. Their relativistic
generalizations may consist of one or two scalar fields,
\cite{qstars-global}, in a Lagrangian with a global $U(1)$
symmetry, or of a non-abelian scalar field in the adjoint
representation of $SU(3)$, \cite{qstars-nonabelian}, or of a
scalar and a fermion field, \cite{qstars-fermion} in
asymptotically flat or anti de Sitter spacetime,
\cite{qstars-AdS}. Q-solitons with local symmetries have also been
investigated. There are charged q-balls, \cite{qballs-charged},
and charged q-stars, \cite{qstars-charged}.

Interesting alternative gravitational theories are the
scalar-tensor gravitational theories, which appeared in the
original paper of Brans and Dicke \cite{bransdicke}, where the
Newtonian constant $G$ was replaced by a scalar field
$\phi_{\textrm{BD}}$, and the total action contained kinetic terms
for the new field times an $\omega_{\textrm{BD}}$ quantity.
$\omega_{\textrm{BD}}$ was regarded as a constant in the original
paper. The theory generalized in a series of papers,
\cite{scalartensor1,scalartensor2}, mainly in the direction of
replacing the constant $\omega_{\textrm{BD}}$ with a function of
the Brans-Dicke (BD) scalar field. Within the BD gravitational
framework, Gunderson and Jensen investigated the coupling of a
scalar field with quartic self-interactions with the metric and
the BD scalar field, $\phi_{\textrm{BD}}$,
\cite{bosonstar-bransdicke1}. The properties of boson stars within
this framework have been extensively studied in a series of papers
\cite{bosonstar-bransdicke2,bosonstar-bransdicke3,bosonstar-bransdicke4}.
Their results generalized in scalar-tensor gravitational theories,
\cite{bosonstar-scalartensor1,bosonstar-scalartensor2}. The case
of charged boson-stars in a scalar-tensor gravitational theory has
been analyzed in \cite{bosonstar-scalartensor-charged}.

The purpose of the present work is to find numerical solutions,
resembling q-stars, of the EYM equations in the presence of a
Higgs doublet in the fundamental representation of $SU(2)$ in BD
theory or in a general scalar-tensor gravitational theory and to
compare the above results with the solutions obtained in General
Relativity. In the absence of the gauge field, the equations of
motion give rise to a gravitating non-topological soliton, when
using a certain potential for which
$\omega_E\equiv\sqrt{U/|\Phi|^2}_{\textrm{min}}<m$, where $m$ is
the mass of the free particles, imposing an harmonic
time-dependence on the scalar field and equalizing the frequency
to $\omega_E$. Our gravitating soliton is \emph{non-topological}
in the sense that $\Phi,U\rightarrow0$ for $\rho\rightarrow\infty$
according to \cite{non-top-sol-review}. It is a \emph{q-type}
non-topological soliton in the sense that in the absence of both
gravitational and gauge fields one can find by simple calculations
that this spherically symmetric Higgs field rotates within its
symmetry space with a frequency $\omega_E$ equal to the minimum of
the $\sqrt{U/|\Phi|^2}$ quantity, as in q-balls. The difference
between this soliton and the usual non-abelian q-balls is that the
symmetry space in the case of non-abelian q-balls is the entire
$SU(3)$ space but in our case is an abelian $U(1)$ subgroup of the
$SU(2)$ group, though both field configurations are non-abelian.
Also, we investigate the fundamental and not the adjoint
representation of $SU(2)$. In any case we find an analytical
solution for the scalar field within the soliton, using the
approximation known by the study of q-stars.

\section{EYM equations in Brans-Dicke gravitational theory}

We consider a Brans-Dicke scalar field $\phi_{\textrm{BD}}$, a
matter Higgs scalar doublet, $\Phi$, in the fundamental
representation of $SU(2)$ and a gauge field $A$, coupled to the
metric $g_{\mu\nu}$. The total action is:
\begin{align}\label{2.1}
S=\frac{1}{16\pi}&\int
d^4x\sqrt{-g}\left(\phi_{\textrm{BD}}R-\omega_{\textrm{BD}}g^{\mu\nu}
\frac{\partial_{\mu}\phi_{\textrm{BD}}\partial_{\nu}\phi_{\textrm{BD}}}{\phi_{\textrm{BD}}}
\right) \nonumber
\\ + &\int d^4x\sqrt{-g}\mathcal{L}_{\textrm{matter}}\ ,
\end{align}
with $\omega_{\textrm{BD}}$ a constant in BD gravity and a certain
function of the $\phi_{\textrm{BD}}$ field in generalized
scalar-tensor gravitational theories and:
\begin{equation}\label{2.2}
\mathcal{L}_{\textrm{matter}}=\frac{1}{4K\textrm{g}^2}\textrm{Tr}F_{\mu\nu}F^{\mu\nu}+
(D_{\mu}\Phi)^{\dagger}(D^{\mu}\Phi)-U\ .
\end{equation}
In the above Lagrangian we define:
\begin{equation}\label{2.3}
\begin{split}
D_{\mu}\Phi&=\partial_{\mu}\Phi-\imath A_{\mu}\Phi \\
F_{\mu\nu}&=\partial_{\mu}A_{\nu}-\partial_{\nu}A_{\mu}-\imath[A_{\mu},A_{\nu}]\
.
\end{split}
\end{equation}
One may use the
$F_{\mu\nu}=\partial_{\mu}A_{\nu}-\partial_{\nu}A_{\mu}-\imath\textrm{g}[A_{\mu},A_{\nu}]$
form and reproduce eq. \ref{2.3} rescaling: $A_{\mu}\rightarrow
A_{\mu}/\textrm{g}$ with $\textrm{g}$ the gauge coupling, or field
strength. The one-form gauge field $A$ is: $A\equiv
A_{\mu}dx^{\mu}\equiv\mathbf{T}_aA^a_{\mu}dx^{\mu}$, with
$\mathbf{T}_a=\frac{1}{2}\tau_a$ and $\tau_a$ the Pauli matrices.
The factor $K$ appearing in the action is defined by the relation
$\textrm{Tr}(\mathbf{T}_a\mathbf{T}_b)=K\delta_{ab}$, reading
$K=1/2$.

In the presence of the Brans-Dicke scalar, the Einstein equations
take the form:
\begin{align}\label{2.4}
G_{\mu\nu}=\frac{8\pi}{\phi_{\textrm{BD}}}T_{\mu\nu}+
\frac{1}{\phi_{\textrm{BD}}}({\phi_{\textrm{BD}}}_{,\mu;\nu}-g_{\mu\nu}
{\phi_{\textrm{BD}}}_{;\lambda}^{\hspace{1em};\lambda} )
\nonumber\\ +
\frac{\omega_{\textrm{BD}}}{\phi_{\textrm{BD}}^2}\left(
\partial_{\mu}\phi_{\textrm{BD}}\partial_{\nu}\phi_{\textrm{BD}}-
\frac{1}{2}g_{\mu\nu}\partial_{\lambda}\phi_{\textrm{BD}}
\partial^{\lambda}\phi_{\textrm{BD}}\right)\ ,
\end{align}
and the equation of motion for the BD field is:
\begin{equation}\label{2.5}
\frac{2\omega_{\textrm{BD}}}{\phi_{\textrm{BD}}}
{\phi_{\textrm{BD}}}_{;\lambda}^{\hspace{1em};\lambda}-
\omega_{\textrm{BD}}
\frac{\partial^{\lambda}\phi_{\textrm{BD}}\partial_{\lambda}\phi_{\textrm{BD}}}
{\phi_{\textrm{BD}}^2}+R=0 \ .
\end{equation}
$G_{\mu\nu}$ is the Einstein tensor, $R$ is the scalar curvature
and $T_{\mu\nu}$ the energy momentum tensor for the matter fields
(gauge and Higgs) given by the equation:
\begin{align}\label{2.6}
T_{\mu\nu}=\frac{2}{\textrm{g}^2}\textrm{Tr}\left(g^{\alpha\beta}F_{\mu\alpha}F_{\nu\beta}
-\frac{1}{4}g_{\mu\nu}F^{\alpha\beta}F_{\alpha\beta}\right)+
\nonumber\\ (D_{\mu}\Phi)^{\dagger}(D_{\nu}
\Phi)+(D_{\mu}\Phi)^T(D_{\nu}\Phi)^{\ast}-g_{\mu\nu}[g^{\alpha\beta}(D_{\alpha}\Phi)^{\dagger}
(D_{\beta}\Phi)]-g_{\mu\nu}U\ .
\end{align}
Tracing Einstein equations and substituting the result in the
Lagrange equation for the BD field we take:
\begin{equation}\label{2.7}
{\phi_{\textrm{BD}}}^{\hspace{1em};\lambda}_{;\lambda}
=\frac{8\pi}{2\omega_{\textrm{BD}}+3}T \ .
\end{equation}

We will choose a general, spherically symmetric field
configuration, defining:
$n^a\equiv(\sin\vartheta\cos\varphi,\sin\vartheta\sin\varphi,\cos\vartheta)$
and $\mathbf{T}_{\rho}=n^a\mathbf{T}_a$,
$\mathbf{T}_{\vartheta}=\partial_{\vartheta}\mathbf{T}_{\rho}$ and
$\mathbf{T}_{\varphi}=(1/\sin\vartheta)\partial_{\varphi}\mathbf{T}_{\rho}$.
The gauge field and the Higgs doublet are:
\begin{align}\label{2.8}
&A=a\mathbf{T}_{\rho}+\imath(1-\textrm{Re}\omega)[\mathbf{T}_{\rho},d\mathbf{T}_{\rho}]+
\textrm{Im}\omega\mathbf{T}_{\rho}= \nonumber\\
&a\mathbf{T}_{\rho}+[\textrm{Im}\omega\mathbf{T}_{\vartheta}+(\textrm{Re}\omega-1)
\mathbf{T}_{\varphi}]d\vartheta+[\textrm{Im}\omega\mathbf{T}_{\varphi}+
(1-\textrm{Re}\omega)\mathbf{T}_{\vartheta}]\sin\vartheta d\varphi
,
\end{align}
\begin{equation}\label{2.9}
\Phi=\sigma\exp (\imath\xi\mathbf{T}_{\rho})|b\rangle
\end{equation}
with $\sigma=\sigma(\rho,t)$, $\xi=\xi(\rho,t)$, $|b\rangle$ a
constant unit vector of the internal $SU(2)$ space of the scalar
(Higgs) field and $a=a_0dt+a_{\rho}d\rho$. In order to form a
field configuration corresponding to a charged q-type soliton we
choose: $a_0=a_0(\rho)$, $\sigma(\rho,t)=\sigma(\rho)$ and
$\xi=\omega_Et$, when we choose $a_{\rho}=0$ for simplicity. The
ansatz $\sigma(\rho,t)=\sigma(r)$ and $\xi=\omega_Et$ is the
obvious generalization to the $\phi(\rho,t)=\sigma(\rho)e^{\imath
\Omega t}$ ansatz, known from q-solitons. So, the role of the
eigen-frequency, $\Omega$ is now played by $\omega_E$. The choice
$a_{\rho}=0$ implies that our field configuration is not very
general, but our purpose is not to find the more general solution,
but a proper one, with the above features and resulting to stable
solitons. The BD scalar field is supposed to be static and to
posses a spherical symmetry, as the matter field configuration.
With the above assumptions, we can write a static, spherically
symmetric metric:
\begin{equation}\label{2.10}
ds^2=-\frac{1}{B}dt^2+\frac{1}{A}d\rho^2+\rho^2d\vartheta^2+\rho^2\sin^2\vartheta
d\varphi^2\ .
\end{equation}
The matter action takes now the form:
\begin{align}\label{2.11}
S_{\textrm{matter}}=\int\frac{\rho^2\sin\vartheta}{\sqrt{AB}}\left[-\frac{1}{2\textrm{g}^2}
\left(a_0'^2AB+2\frac{|\omega|^2a_0^2}{\rho^2}+\frac{(|\omega|^2-1)^2}{\rho^4}\right)
+\sigma'^2A \right. \nonumber\\ \left.
-\frac{1}{4}(\omega_E-a_0)^2\sigma^2B+\frac{\sigma^2}{2\rho^2}
[(\textrm{Re}\omega-\cos(\omega_Et))^2+(\textrm{Im}\omega-\sin(\omega_Et))^2]-U
\right]
\end{align}
In order the action to be time-independent we may choose
$\textrm{Re}\omega=\cos(\omega_Et)$ and
$\textrm{Im}\omega=\sin(\omega_Et)$, but this choice is not a
solution to the equation of motion for $\omega$, or $\omega=0$
which is a solution of the equation of motion, so, our solution is
embedded abelian.

If $m$ is the mass of the free particles, we make the following
rescalings:
\begin{align}\label{2.12}
\tilde{\rho}=2m\rho \ ,\hspace{1em}
\tilde{\omega}_E=\frac{\omega_E}{2m} \ ,\hspace{1em}
\tilde{a}_0=\frac{a_0}{2m} \ ,\hspace{1em}
\tilde{\sigma}=\frac{\sigma}{\frac{m}{2}}\ ,\nonumber\\
\tilde{r}=\epsilon\tilde{\rho}\
,\hspace{1em}\tilde{\textrm{g}}=\textrm{g}\epsilon^{-1}\
,\hspace{1em}
\Phi_{\textrm{BD}}=\frac{2\omega_{\textrm{BD}}+4}{2\omega_{\textrm{BD}}+3}G\phi_{\textrm{BD}}\
,
\end{align}
with:
\begin{equation}\label{2.13}
\epsilon\equiv\sqrt{8\pi Gm^2}\ .
\end{equation}
We define:
\begin{equation}\label{2.14}
W=\left(\frac{d\Phi}{dt}\right)^{\dagger}\left(\frac{d\Phi}{dt}\right)\
,\hspace{1em}V=\left(\frac{d\Phi}{d\rho}\right)^{\dagger}\left(\frac{d\Phi}{d\rho}\right)\
.
\end{equation}
Gravity becomes important when $R\sim8\pi GM$, with $<\phi>=1/G$.
With our rescalings and because the energy density within the
soliton is $\sim m^4$, we find that $\tilde{r}\sim1$ and if
$\sigma$ varies very slowly within the soliton, from a $\sigma(0)$
value at $\tilde{r}=0$ to a zero value at the outer edge of the
soliton surface, then $V\sim\epsilon^2m^4$. For $m\sim GeV$ the
$O(\epsilon)$ quantities are negligible. We choose a simple
rescaled potential, admitting q-ball type solutions in the absence
of gravity and gauge fields, namely:
\begin{equation}\label{2.15}
U=m^2\Phi^{\dagger}\Phi\left(1- \frac{4}{m^2}
\Phi^{\dagger}\Phi+\frac{16}{3m^4}(\Phi^{\dagger}\Phi)^2\right)\ ,
\end{equation}
which with our rescalings and after some algebra takes the form:
\begin{equation}\label{2.16}
\widetilde{U}=\frac{\tilde{\sigma}^2}{4}\left(1-\tilde{\sigma}^2
+\frac{\tilde{\sigma}^4}{3}\right)\ ,
\end{equation}
where we set $m=1$. From now on we drop the tildes and the
$O(\epsilon)$ quantities. From the equation of motion for the
Higgs field we find:
\begin{equation}\label{2.17}
\begin{split}
\sigma^2=1+\theta_0B^{1/2}\
,\hspace{1em}U&=\frac{1}{12}(1+\theta_0^3B^{3/2})\
,\hspace{1em}W=\theta_0^2B(1+\theta_0B^{1/2})\ ,\\ T&=2W-4U\ ,
\end{split}
\end{equation}
with:
\begin{equation}\label{2.18}
\theta_0=\omega_E-a_0\ .
\end{equation}

The equation of motion for the Higgs field within the surface
gives a boundary condition for the gauge field $\theta_0$, which
reduces to an eigenvalue equation for the frequency in the case of
global $SU(2)$ symmetry (i.e.: when $a_0=0$). The surface width is
of $O(m^{-1})$. The Higgs field $\sigma$ varies rapidly from a
$\sigma_0$ value at the inner edge of the surface, to a zero value
at the outer one. Dropping form the Lagrange equation the
$O(\epsilon)$ terms and integrating the resulting equation, we
find that within the surface:
\begin{equation}\label{2.19}
V+W-U=0\ .
\end{equation}
In order to match the interior with the surface solution we set
$\sigma'=0$ at the inner edge of the surface. Then, using eqs.
\ref{2.17} and \ref{2.19} we find:
\begin{equation}\label{2.20}
{\theta_0}_{\textrm{sur}}=\frac{A^{1/2}_{\textrm{sur}}}{2}=\frac{B_{\textrm{sur}}^{-1/2}}{2}\
,
\end{equation}
where ${\theta_0}_{\textrm{sur}}$ is the value of $\theta_0$
within the thin surface. In the absence of gauge fields we take:
$\omega_E=A_{\textrm{sur}}^{1/2}/2$ which in the absence of
gravity gives $\omega_E=1/2$, which is the correct eigenvalue
equation for the q-soliton frequency.

With the above definitions, the independent Einstein equations
take the following form:
\begin{align}\label{2.21}
\frac{A-1}{r^2}+\frac{1}{r}\frac{dA}{dr}=\frac{2\omega_{\textrm{BD}}+3}
{(2\omega_{\textrm{BD}}+4)\Phi_{\textrm{BD}}}
\left(-W-U-\frac{2W-4U}{2\omega_{\textrm{BD}}+3}\right) \nonumber
\\ -\frac{\omega_{\textrm{BD}}A}{2\Phi_{\textrm{BD}}^2}
\left(\frac{d\Phi_{\textrm{BD}}}{dr}\right)^2-
\frac{A}{2\Phi_{\textrm{BD}}B}\frac{dB}{dr}\frac{d\Phi_{\textrm{BD}}}{dr}\
,
\end{align}
\begin{align}\label{2.22}
\frac{A-1}{r^2}-\frac{A}{B}\frac{1}{r}\frac{dB}{dr}=
\frac{2\omega_{\textrm{BD}}+3}{(2\omega_{\textrm{BD}}+4)\Phi_{\textrm{BD}}}
\left(W-U-\frac{2W-4U}{2\omega_{\textrm{BD}}+3}\right) \nonumber
\\ +\frac{\omega_{\textrm{BD}}A}{2\Phi_{\textrm{BD}}^2}
\left(\frac{d\Phi_{\textrm{BD}}}{dr}\right)^2
+\frac{A}{\Phi_{\textrm{BD}}}\left(\frac{d^2\Phi_{\textrm{BD}}}{dr^2}+
\frac{1}{2A}\frac{dA}{dr}\frac{d\Phi_{\textrm{BD}}}{dr}\right)\ ,
\end{align}
the Euler-Lagrange equation for the BD scalar is:
\begin{equation}\label{2.23}
A\left[\frac{d^2\Phi_{\textrm{BD}}}{dr^2}+
\left(\frac{2}{r}+\frac{1}{2A}\frac{dA}{dr}-\frac{1}{2B}\frac{dB}{dr}\right)
\frac{d\Phi_{\textrm{BD}}}{dr}\right]=
\frac{2W-4U}{2\omega_{\textrm{BD}}+4}\ ,
\end{equation}
and the equation of motion for the new gauge field $\theta_0$:
\begin{equation}\label{2.24}
\theta_0''+\left(\frac{2}{r}+\frac{A'}{2A}+\frac{B'}{2B}\right)\theta_0'-\frac{\textrm{g}^2
\theta_0(1+\theta_0B^{1/2})}{2A}=0\ ,
\end{equation}
with boundary conditions:
\begin{equation}\label{2.25}
A(0)=1\ ,\hspace{1em} \theta'(0)=0\
,\hspace{1em}\theta(R)=\theta_{\textrm{sur}}\
,\hspace{1em}\Phi'_{\textrm{BD}}=1\ ,
\end{equation}
and for $r\rightarrow\infty$:
\begin{equation}\label{2.26}
A(r)=1/B(r)=1\ ,\hspace{1em}\Phi_{\textrm{BD}}=1\
,\hspace{1em}\theta(r)=\omega_E\ .
\end{equation}
The first condition of eq. \ref{2.24} reflects our freedom to
redefine $A(r)$, the second and forth result from the spherical
symmetry of the configuration and the third is the eigenvalue
equation for the new gauge field $\theta_0$, with $R$ the soliton
radius. The first condition of eq. \ref{2.25} is a straightforward
consequence of the Einstein equations for localized matter
configurations, the second is the boundary condition for the BD
field with the proper rescalings and the third denotes the absence
of gauge fields at infinity. We numerically solve the coupled
system of eqs. \ref{2.21}-\ref{2.24}.

\begin{figure}
\centering
\includegraphics{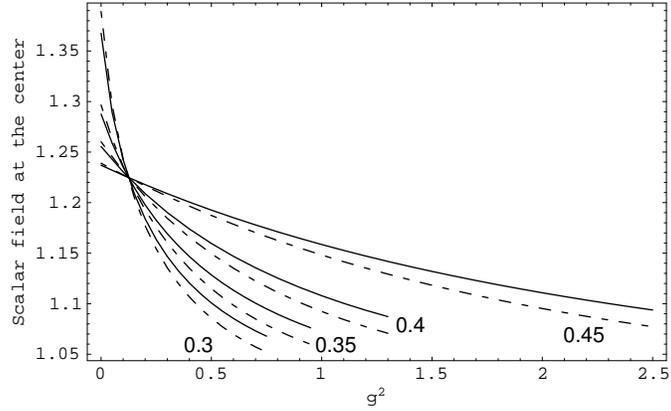}
\caption{The value of the Higgs field $\sigma$ at the center of
the soliton as a function of the coupling constant $g^2$. The
numbers within the figures \ref{figure1}-\ref{figure5} denote the
eigenvalue ${\theta_0}_{\textrm{sur}}$, which for $\textrm{g}^2=0$
reduces to $\omega_E$. Dashed lines correspond to
$\omega_{\textrm{BD}}=5$ and solid lines to
$\omega_{\textrm{BD}}=500$. The results of General Relativity
almost coincide with the BD theory for
$\omega_{\textrm{BD}}=500$.} \label{figure1}
\end{figure}

\begin{figure}
\centering
\includegraphics{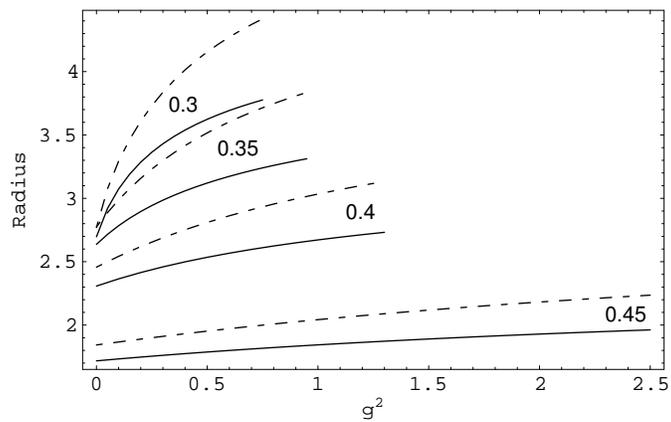}
\caption{The radius of the soliton as a function of $g^2$.}
\label{figure2}
\end{figure}

\begin{figure}
\centering
\includegraphics{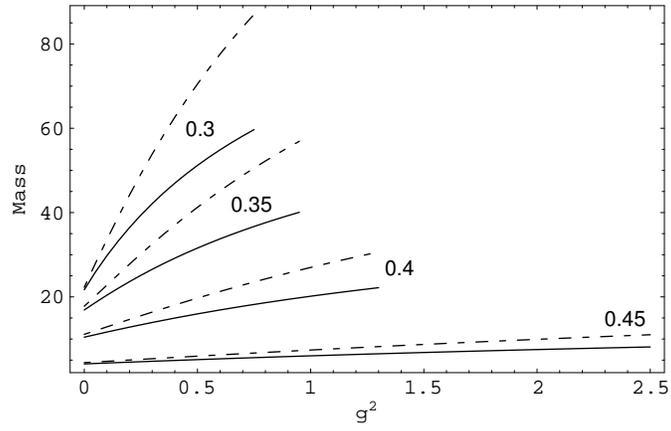}
\caption{The asymptotically anti de Sitter mass, $M$, as a
function of $g^2$.} \label{figure3}
\end{figure}

\begin{figure}
\centering
\includegraphics{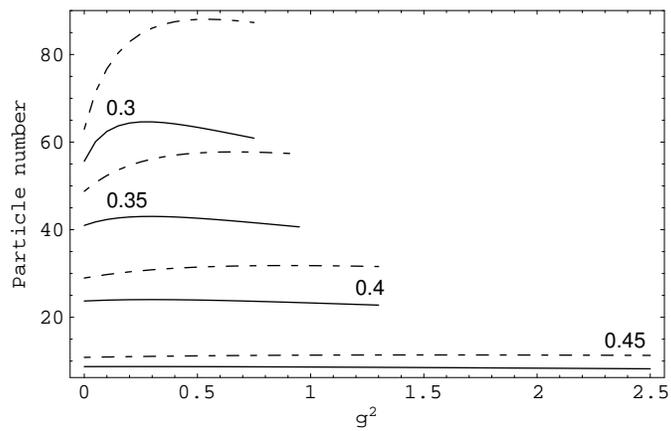}
\caption{The particle number, $N$, of the soliton as a function of
$g^2$.} \label{figure4}
\end{figure}

\begin{figure}
\centering
\includegraphics{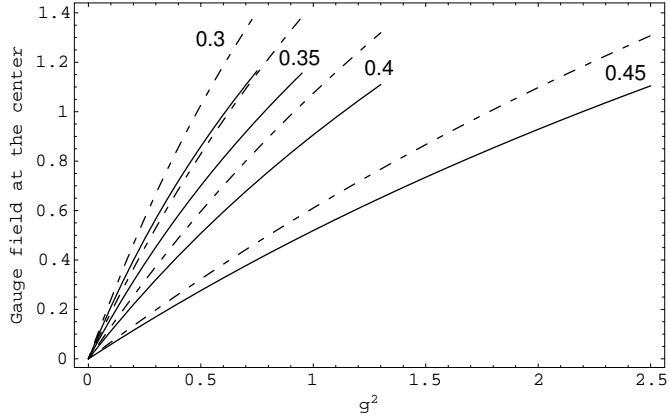}
\caption{The value of the gauge field, $a_0$, at the center of the
soliton as a function of $g^2$.} \label{figure5}
\end{figure}

The Noether currents corresponding to the generators of the
$SU(2)$ algebra are given by the relation:
\begin{equation}\label{2.27}
j_{0\alpha}=\left(\begin{array}{cc}
  \frac{\partial L}{\partial(\partial_0\Phi)} & \frac{\partial
  L}{\partial(\partial_0\Phi^{\ast})}
\end{array}\right)\left(\begin{array}{cc}
  \imath T_{\alpha} & 0 \\
  0 & -\imath T_{\alpha}
\end{array}\right)\left(\begin{array}{c}
  \Phi\ \\
  \Phi^{\ast}
\end{array}\right)\ .
\end{equation}
We can find that:
\begin{equation}\label{2.28}
\begin{split}
j_{01}&=\frac{1}{2}\sigma^2\theta_0\sin\vartheta\cos\varphi\ , \\
j_{02}&=\frac{1}{2}\sigma^2\theta_0\sin\vartheta\sin\varphi\ , \\
j_{03}&=\frac{1}{2}\sigma^2\theta_0\cos\vartheta\ ,
\end{split}
\end{equation}
and
\begin{equation}\label{2.29}
j_0\equiv\sqrt{j_{01}^2+j_{02}^2+j_{03}^2}=\frac{1}{2}\sigma^2\theta_0\
.
\end{equation}
The particle number is:
\begin{equation}\label{2.30}
N=2\pi\int\sigma^2\theta_0\sqrt{\frac{A}{B}}r^2dr\ .
\end{equation}
The total energy of the field configuration results from the
relation:
\begin{equation}\label{2.31}
A(\rho)=1-\frac{2GM}{\rho}+\frac{G\textrm{g}^2N^2}{4\pi\rho^2}\
,\hspace{1em}\rho\rightarrow\infty\ ,
\end{equation}
which gives with our rescalings:
\begin{equation}\label{2.32}
M=4\pi r\left(1-A(r)+\frac{\textrm{g}^2N^2}{32\pi^2r^2}\right)\
,\hspace{1em}r\rightarrow\infty\ .
\end{equation}

For $\omega_{\textrm{BD}}\rightarrow\infty$ the Einstein equations
take the simple form:
\begin{equation}\label{2.33}
\frac{A-1}{r^2}+\frac{A'}{r}=-U-W-\frac{\theta_0'^2}{2\textrm{g}^2}AB\
,
\end{equation}
\begin{equation}\label{2.34}
\frac{A-1}{r^2}-\frac{A}{B}\frac{B'}{r}=W-U-\frac{\theta_0'^2}{2\textrm{g}^2}AB\
,
\end{equation}
when the other relations remain the same and the results of
General Relativity are reproduced.

\section{EYM equations in general scalar-tensor theory}

In the original BD gravitational theory $\omega_{\textrm{BD}}$ is
a constant. In a more general theory it may be regarded as a
function, usually of the BD field. We will use one of the forms
investigated in a cosmological framework,
\cite{scalartensor-cosmological,scalartensor-cosm-analytical},
namely:
\begin{equation}\label{3.1}
2\omega_{\textrm{BD}}+3=\omega_0\phi_{\textrm{BD}}^n\ ,
\end{equation}
with $\omega_0$ and $n$ constants. The Lagrange equation for the
BD field is:
\begin{equation}\label{3.2}
{\phi_{\textrm{BD}}}^{\hspace{1em};\lambda}_{;\lambda}=\frac{1}{\omega_0\phi_{\textrm{BD}}^n}
\left(8\pi T-\frac{d\omega_{\textrm{BD}}}{d\phi_{\textrm{BD}}}
{\phi_{\textrm{BD}}}^{,\rho}{\phi_{\textrm{BD}}}_{,\rho}\right)\ ,
\end{equation}

\begin{figure}
\centering
\includegraphics{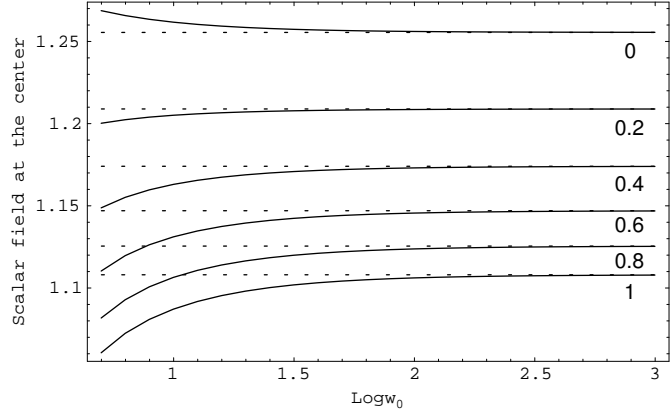}
\caption{The value of the field $\sigma$ at the center of the
soliton as a function of $\omega_0$. The numbers within the
figures \ref{figure6}-\ref{figure10} denote the $\textrm{g}^2$
value. We use ${\theta_0}_{\textrm{sur}}=0.4$. In figures
\ref{figure6}-\ref{figure10} solid lines are the numerical results
produced in the framework of scalar-tensor gravitational theory,
when dashed lines are the results from General Relativity}
\label{figure6}
\end{figure}

\begin{figure}
\centering
\includegraphics{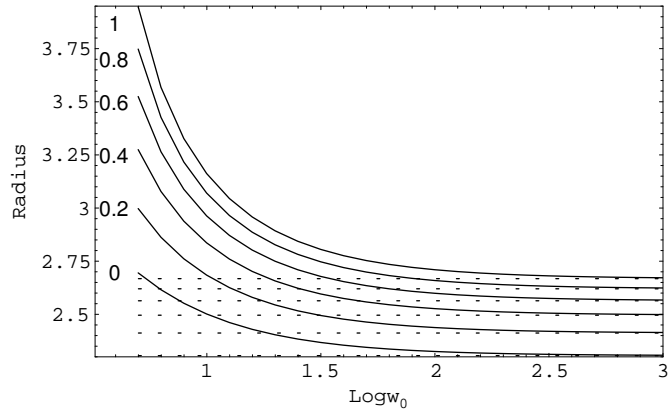}
\caption{The radius of the soliton as a function of $\omega_0$}
\label{figure7}
\end{figure}

\begin{figure}
\centering
\includegraphics{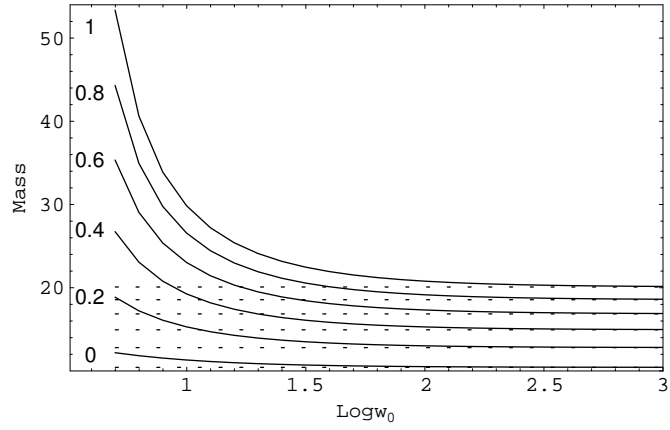}
\caption{The asymptotically anti de Sitter mass, $M$ of the
soliton as a function of $\omega_0$.} \label{figure8}
\end{figure}

\begin{figure}
\centering
\includegraphics{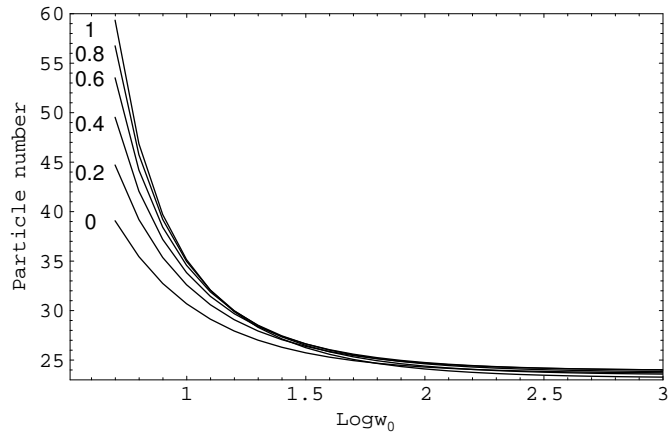}
\caption{The particle number of the soliton as a function of
$\omega_0$.} \label{figure9}
\end{figure}

\begin{figure}
\centering
\includegraphics{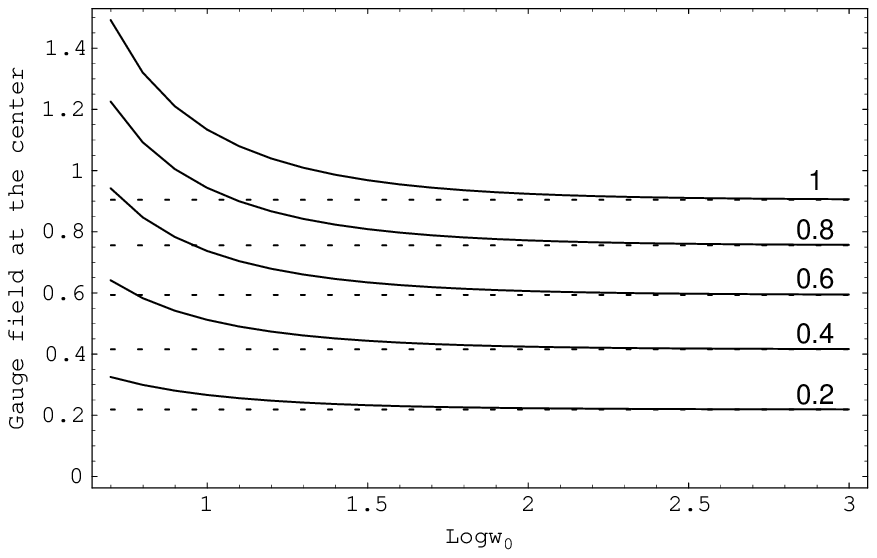}
\caption{The value of the gauge field at the center of the soliton
as a function of $\omega_0$.} \label{figure10}
\end{figure}

We rescale:
\begin{equation}\label{3.3}
\tilde{\omega}_0=\left(\frac{2\omega_{\textrm{BD}}+3}{2\omega_{\textrm{BD}}+4}\right)^n
G^n\omega_0\ ,
\end{equation}
and the other quantities as in \ref{2.12} and drop the tildes and
the $O(\epsilon)$ quantities. We will use $n=1$. For other values
of $n$ the behavior of the soliton parameters is very similar. The
Einstein equations take the following form:
\begin{align}\label{3.4}
\frac{A-1}{r^2}+\frac{1}{r}\frac{dA}{dr}=
\frac{\omega_0}{\omega_0\Phi_{\textrm{BD}}+1}\left[-W-U-\frac{1}
{\omega_0\Phi_{\textrm{BD}}} \right.\times\nonumber\\ \left.
\left(2W-4U-\frac{A\Phi_{\textrm{BD}}'^2}{2}
\frac{\omega_0\Phi_{\textrm{BD}}+1}{\Phi_{\textrm{BD}}}\right)\right]
\nonumber\\
-\frac{\omega_0\Phi_{\textrm{BD}}-3}{2}\frac{A\Phi_{\textrm{BD}}'^2}{2\Phi_{\textrm{BD}}^2}
-\frac{AB'\Phi_{\textrm{BD}}'}{2\Phi_{\textrm{BD}}B}\ ,
\end{align}
\begin{align}\label{3.5}
\frac{A-1}{r^2}-\frac{A}{B}\frac{1}{r}\frac{dB}{dr}=
\frac{\omega_0}{\omega_0\Phi_{\textrm{BD}}+1}\left[W-U-\frac{1}
{\omega_0\Phi_{\textrm{BD}}} \right.\times\nonumber\\ \left.
\left(2W-4U-\frac{A\Phi_{\textrm{BD}}'^2}{2}
\frac{\omega_0\Phi_{\textrm{BD}}+1}{\Phi_{\textrm{BD}}}\right)\right]
\nonumber\\+\frac{\omega_0\Phi_{\textrm{BD}}-3}{2}\frac{A\Phi_{\textrm{BD}}'^2}
{2\Phi_{\textrm{BD}}^2}+\frac{A\Phi_{\textrm{BD}}''}{\Phi_{\textrm{BD}}}+
\frac{A'\Phi_{\textrm{BD}}'}{2\Phi_{\textrm{BD}}}\ .
\end{align}
The equation of motion for the BD field is:
\begin{align}\label{3.6}
A\left[\Phi_{\textrm{BD}}''+\left(\frac{2}{r}+\frac{A'}{2A}-\frac{B'}{2B}\right)
\Phi_{\textrm{BD}}'\right]=\nonumber\\
\frac{1}{\omega_0\Phi_{\textrm{BD}}+1}
\left[2W-4U-\frac{A\Phi_{\textrm{BD}}'^2}{2}\frac{\omega_0\Phi_{\textrm{BD}}+1}
{\Phi_{\textrm{BD}}}\right]\ .
\end{align}
The equations of motion for the gauge and Higgs field remain
unchanged. We solve numerically the coupled system of eqs.
\ref{3.3}-\ref{3.6} and \ref{2.24}, when the relations \ref{2.30},
\ref{2.32} for the soliton mass and particle number hold true.

\section{Conclusions}

We studied EYM equations in the presence of a Higgs doublet in
Brans-Dicke and a simple scalar-tensor gravitational theory and
compared our results with the solutions of General Relativity. The
Higgs doublet is characterized by a potential admitting q-star
type and and q-ball type solutions in the absence of gauge fields
and gravity and gauge fields respectively. It is a matter of
simple algebra to verify this claim. So, the EYM-Higgs equations
\emph{reduce} to a system of equations corresponding to (charged)
soliton stars. These objects are \emph{stars} and not black holes,
having no horizon or other anomalies.

There are two crucial parameters, resulting from the soliton star
itself, the field strength, $\textrm{g}$ and the eigenvalue
${\theta_0}_{\textrm{sur}}$ which in the absence of gauge fields
reduces to the usual soliton eigen-frequency. The above eigenvalue
is straightforward connected to gravity strength on the surface,
through eq. \ref{2.20}. So, a soliton with small
${\theta_0}_{\textrm{sur}}$ shows a stronger gravitational force
on its surface, which corresponds to a more massive, or denser
soliton, and this can be verified by our figures. Also, larger
value for the field strength, increases the energy and radius of
the field configuration due to the electrostatic repulsion between
the different parts of the soliton, when the value of the scalar
field within the soliton decreases for the same reason. These
results hold true in General Relativity as well as in
scalar-tensor theories.

In figures \ref{figure1}-\ref{figure5} the results of General
Relativity are not depicted because they almost coincide with the
$\omega_{\textrm{BD}}=500$ (solid lines) case.
$\omega_{\textrm{BD}}\simeq500$ is the lower experimental limit.
The results of general relativity are exactly reproduced when
$\omega{\textrm{BD}}\rightarrow\infty$. In figures
\ref{figure6}-\ref{figure10} we study the behavior of the soliton
parameters for $5\leq\omega_0\leq1000$. The results of General
Relativity are practically reproduced for $\omega_0=1000$ as one
can see from the dashed lines in figures
\ref{figure6}-\ref{figure10}, which correspond to the results of
General Relativity.

\vspace{1em}

\textbf{ACKNOWLEDGEMENTS}

\vspace{1em}

I wish to thank N. D. Tracas, E. Papantonopoulos and P.
Manousselis for helpful discussions.

\end{document}